\documentclass[twocolumn,superscriptaddress,longbibliography]{revtex4-1}
\usepackage{graphicx}
\usepackage{times}
\usepackage{amsmath}
\usepackage{amssymb}
\usepackage{units}
\usepackage{stmaryrd} 
\usepackage[compact]{titlesec}
\titlespacing{\section}{0pt}{*0}{*0}
\titlespacing{\subsection}{0pt}{*0}{*0}
\titlespacing{\subsubsection}{0pt}{*0}{*0}

\begin{document}
\preprint{0}

\title{Kondo hybridization and the origin of metallic states at the (001) surface of SmB$_{6}$}

\author{E. Frantzeskakis}
\email{e.frantzeskakis@uva.nl} 
\address{Van der Waals - Zeeman Institute, University of Amsterdam, Science Park 904, 1098 XH, Amsterdam, the Netherlands}

\author{N. de Jong}
\address{Van der Waals - Zeeman Institute, University of Amsterdam, Science Park 904, 1098 XH, Amsterdam, the Netherlands}

\author{B. Zwartsenberg}
\address{Van der Waals - Zeeman Institute, University of Amsterdam, Science Park 904, 1098 XH, Amsterdam, the Netherlands}

\author{Y. K. Huang}
\address{Van der Waals - Zeeman Institute, University of Amsterdam, Science Park 904, 1098 XH, Amsterdam, the Netherlands}

\author{Y. Pan}
\address{Van der Waals - Zeeman Institute, University of Amsterdam, Science Park 904, 1098 XH, Amsterdam, the Netherlands}

\author{X. Zhang}
\address{Key Laboratory of Advanced Functional Materials, Ministry of Education, College of Materials Science and Engineering, Beijing University of Technology, Beijing 100124, China}

\author{J. X. Zhang}
\address{Key Laboratory of Advanced Functional Materials, Ministry of Education, College of Materials Science and Engineering, Beijing University of Technology, Beijing 100124, China}

\author{F. X. Zhang}
\address{Key Laboratory of Advanced Functional Materials, Ministry of Education, College of Materials Science and Engineering, Beijing University of Technology, Beijing 100124, China}

\author{L. H. Bao}
\address{Inner Mongolia Key Laboratory for Physics and Chemistry of Functional Materials, Inner Mongolia Normal University, Hohhot 010022, China}

\author{O. Tegus}
\address{Inner Mongolia Key Laboratory for Physics and Chemistry of Functional Materials, Inner Mongolia Normal University, Hohhot 010022, China}

\author{A. Varykhalov}
\address{Helmholtz-Zentrum Berlin f\"{u}r Materialien und Energie, Albert-Einstein-Strasse 15, 12489 Berlin, Germany}

\author{A. de Visser}
\address{Van der Waals - Zeeman Institute, University of Amsterdam, Science Park 904, 1098 XH, Amsterdam, the Netherlands}

\author{M. S. Golden}
\email{m.s.golden@uva.nl} 
\address{Van der Waals - Zeeman Institute, University of Amsterdam, Science Park 904, 1098 XH, Amsterdam, the Netherlands}


\begin{abstract}
SmB$_{6}$, a well-known Kondo insulator, has been proposed to be an ideal topological insulator with states of topological character located in a clean, bulk electronic gap, namely the Kondo hybridization gap.
Seeing as the Kondo gap arises from many body electronic correlations, this would place SmB$_{6}$ at the head of a new material class: topological Kondo insulators. Here, for the first time, we show that the \textit{k}-space characteristics of the Kondo hybridization process is the key to unravelling the origin of the two types of metallic states experimentally observed by ARPES in the electronic band structure of SmB$_{6}$(001).
One group of these states is essentially of bulk origin, and cuts the Fermi level due to the position of the chemical potential 20 meV above the lowest lying 5$d$-4$f$ hybridization zone. 
The other metallic state is more enigmatic, being weak in intensity, but represents a good candidate for a topological surface state.
However, before this claim can be substantiated by an unequivocal measurement of its massless dispersion relation, our data raises the bar in terms of the ARPES resolution required, as we show there to be a strong renormalization of the hybridization gaps by a factor 2-3 compared to theory, following from the knowledge of the true position of the chemical potential and a careful comparison with the predictions from recent LDA+Gutzwiller calculations.
All in all, these key pieces of evidence act as triangulation markers, providing a detailed description of the electronic landscape in SmB$_{6}$, pointing the way for future, ultrahigh resolution ARPES experiments to achieve a direct measurement of the Dirac cones in the first topological Kondo insulator.
\end{abstract}

\maketitle

\section*{Introduction}

Kondo insulators (KI) have attracted enormous scientific interest as they present a vivid manifestation of strong electronic correlations that give rise to exotic ground states in real materials \cite{Riseborough2000, Takabatake1998, Aeppli1992, Tsunetsugu1997}.
They are based on elements with a partially-filled $f$-shell. 
In such systems, the magnetic moment of localized $f$-electrons can be effectively screened by delocalized conduction carriers (i.e. $c$-$f$ interaction) leading to a hybridization gap opening: the fingerprint of Kondo behavior in the electronic band structure.
In many respects SmB$_{6}$ is a typical KI \cite{Menth1969, Souma2002, Nozawa2002, Mizumaki2009, Kasuya1994}, except that its residual resistivity at low temperatures is incompatible with the Kondo scenario \cite{Cooley1995, Allen1979, Derr2008, Nickerson1971, Nanba1993, Curnoe2000}. 

Interest in SmB$_{6}$ has been recently rekindled since band structure calculations \cite{Dzero2010, Alexandrov2013, Lu2013} and transport
measurements \cite{Botimer2013, Kim2013, Li2013, Wolgast2013, Zhang2013} have led to the proposition that electronic states of topological character might exist within the hybridization gap, and be responsible for the non-divergent behavior of the resistivity at low temperatures.
If this scenario is correct, not only is the long-standing challenge of understanding SmB$_{6}$'s behavior at low temperature solved, but also the first member of a new class of novel states of matter - topological Kondo insulators (TKI) \cite{Dzero2010} - has been discovered.
More than just representing a further 'stamp' in the growing collection of novel quantum states of (electronic) matter, SmB$_{6}$ could represent a very promising alternative to the Bi-based 3D topological insulators (TI's). 
In theory, these are systems with a bulk electronic gap at the Fermi level and conducting surface states, topologically protected from back-scattering through a chiral spin texture \cite{Hasan2010}. 
Such compounds are promising for a wide spectrum of applications ranging from spintronics \cite{Bianchi2010,King2011,Bahramy2011} to quantum computation \cite{Nayak2008}. 
However, in real crystals of these Bi-based (and related) 3D TI's, the effect of topologically trivial, transport active defect states in the bulk gap \cite{Analytis2010} often mask the signatures of the topological surface states in transport experiments, and at present this situation effectively prevents their exploitation in devices.

First, encouraging signs are emerging of charge transport involving topologically non-trivial states in SmB$_{6}$ \cite{Li2013, Kim2013}, making further experimental investigation of the electronic structure of this system imperative.

As the hallmark of (strong) topological behavior is the observation of an odd number of Dirac cones in the electronic band structure (and their special spin texture), angle resolved photoelectron spectroscopy (ARPES) has played a central role in the experimental investigation of 3D TI's to date \cite{Hasan2010}.
Intriguingly, recent ARPES studies on the (001) cleavage surface of SmB$_{6}$ crystals \cite{Miyazaki2012, Neupane2013, Jiang2013, Xu2013} have reported metallic states but failed to clearly identify a Dirac cone, thus leaving the question of the topological origin of the observed states unanswered.

\begin{figure*}
  \centering
  \includegraphics[width = 16 cm]{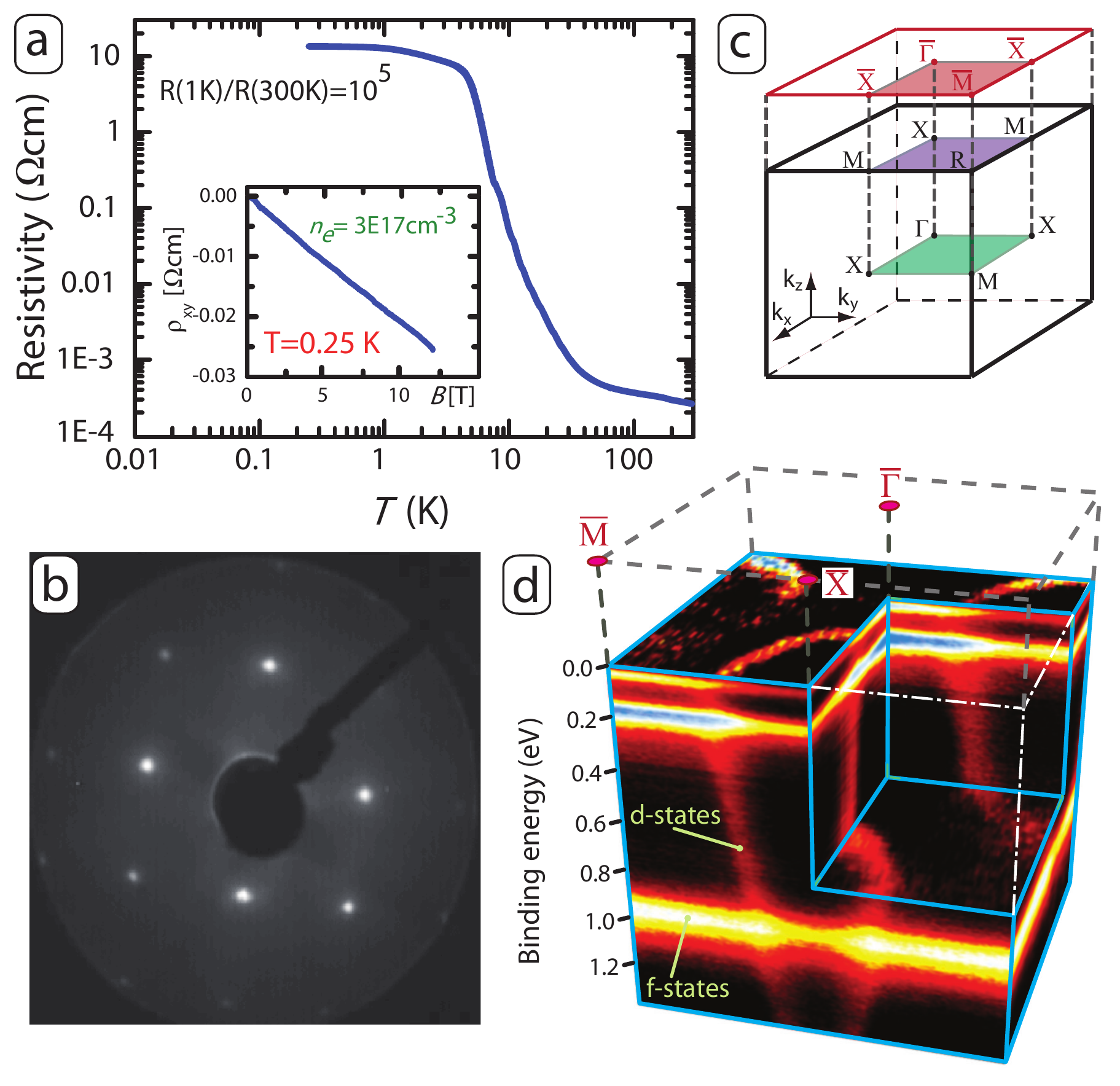}
  \caption{
\textbf{SmB$_{6}$ transport, surface quality and ARPES data overview}
(a) Temperature dependence of the resistivity of a floating-zone, optical-furnace-grown SmB$_{6}$ crystal. There is a sharp increase below 50K which results in a very high resistivity ratio of 10$^{5}$ between the saturation regime and room temperature. The inset shows the Hall resistivity, which is linear, yielding an n-type carrier concentration as indicated (low T regime).
(b) Very sharp low-energy electron diffraction (LEED) spots reflect the simple (1x1) surface structure - free of reconstructions - and attest to the long-range order of the cleavage surface. The LEED primary beam energy was 97 eV. 
(c) Sketch of the 3D Brillouin zone (BZ) of SmB$_{6}$ and its projection onto the (001) plane. The colored planes denote the relevant cuts through the zone that would be made for photon energies in ARPES corresponding to a $k_{\textmd{z}}$ value of an even number of $\pi$/a (green); an odd number of $\pi$/a (purple) and the surface projection (red).
(d) Three-dimensional ARPES data-block displaying $I$($E_{\textmd{B}},k_{\textmd{x}}$,$k_{\textmd{y}}$). Two types of bands are seen with the strongly dispersing(flat) bands originating from Sm 5$d$(4$f$) states, as indicated. The Sm 5$d$-related states give rise to elliptical contours at higher binding energies (the cutaway in the lower right corner of the data-block showing this for a binding energy of 0.8 eV), which then evolve to give elliptical Fermi surface contours around the $\overline{\textmd{X}}$ points. These states are discussed in the text as the $\textmd{X}$-states, and are argued to be essentially bulk-derived bands, well described by bulk, \textit{ab initio} theoretical approaches. Where the flat 4$f$-bands intersect the 5$d$-dominated $\textmd{X}$-states, complex band structures result due to the hybridization effects between $d$ and $f$ states of the equal symmetry, as will be discussed in detail in the context of Figure 2.   
The photon energy for the data-block was $h\nu=70$ eV.
}
\label{SmB6_Fig1_v5}
\end{figure*}

Here we present clear experimental evidence from ARPES on high-quality floating-zone grown SmB$_{6}$ crystals for two different metallic states in this system, and we provide a self-consistent and robust explanation of the origin of both these groups of metallic states, which rests on the cornerstone of the Kondo insulator behavior of SmB$_{6}$: the $d$-$f$ hybridization.
Our ARPES data provide three vital triangulation points for understanding the electronic structure of SmB$_{6}$ crystals.
Firstly, a complete determination of the bulk 5$d$-related bands, the 4$f$-related features and their hybridization at energy scales from the order of 1 eV to a few 10's of meV is provided.
Secondly and thirdly, within this framework we succeed in an unambiguous identification of both the position of the chemical potential and the magnitude of the hybridization gaps in the low-lying electronic structure.
Importantly, we discover that both of these experimentally determined characteristics of the near-surface electronic states of SmB$_{6}$ are - as yet - incorrectly described by theory.

Combining the three key experimental ingredients from our ARPES data, we are able to pin-point the exact locations in the $E(\bf{k})$ landscape of SmB$_{6}$ in which very small hybridization gaps of 5-10 meV occur. It is in these gaps - which are \textit{not} located at the Fermi energy of cleaved SmB$_{6}$ - in which ultrahigh resolution ARPES investigations will have to search for the elusive Dirac cones which will prove the topological character of the Kondo insulating behavior in this material. 
It is clear that the insight won here generates a well-defined and stringent framework, in which all future \textit{k}-sensitive investigations of SmB$_{6}$ will take their place.\\       

\begin{figure}
  \centering
  \includegraphics[width = 6.0 cm]{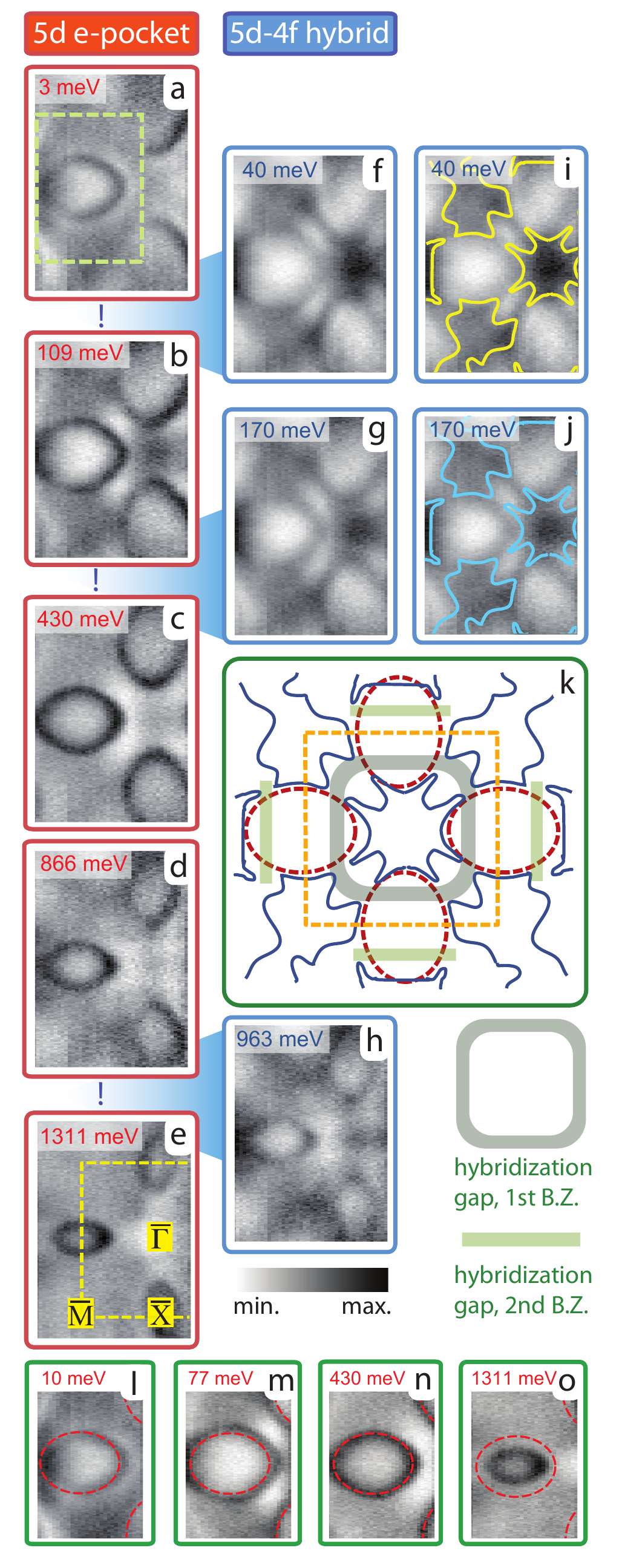}
  \caption{
\textbf{\textit{k}-space fingerprint of 5$d$-4$f$ hybridization}
(a)-(e): The elliptical contours from Sm 5$d$-dominated states centered at $\overline{\textmd{X}}$ increase in size as the binding energy decreases. The yellow dashed square in (e) shows the surface Brillouin zone.
Panels (f)-(h) illustrate that when the $d$ and $f$ states interact strongly - in hybridization zones - strong spectral weight redistribution occurs, giving significant intensity at $\overline{\Gamma}$.
In (i) the constant energy contours within the lowest lying hybridization zone are indicated in yellow. Panel (j) shows these same lines (now in blue) to also be an excellent fit to the energy contours of the deeper lying hybridization zone.  
Panel (k) laysover exemplary hybridized (blue) and non-hybridized (red-dashed) contours. The broad, green-shaded lines highlight the resulting fourfold symmetric \textit{k}-space fingerprint of the 5$d$-4$f$ hybridization, characterized by a gapping-out of the ARPES intensity underlying parts of the red-dashed ellipses.
(l)-(o) show zooms of data covering the $k$-space area indicated with a green dashed box in panel (a).
In (n), red-dashed ellipses are drawn to match the data at 430 meV binding energy and the same ellipses are copied onto the other top-most panels, reiterating the trend also evident in panels (a)-(e) and in Figure 1d that the $\textmd{X}$-state ellipses grow in size as the binding energy is increased. Panel (l) highlights that just after a hybridzation zone, the $\textmd{X}$-state ellipses shrink again, which leads to a \textit{k}-discontinuity for these states. This second characteristic sign of strong 5$d$-4$f$ interaction is also shown in Figure 3 and discussed in detail in the context of Figure 4.
The ARPES data were acquired with $h\nu=70$ eV.
}
\label{SmB6_Fig2_v5}
\end{figure}

\section*{Results and Discussion}

The transport data shown in Figure 1a attest to the very high quality of the flux-free single crystals studied. The resistivity displays a sharp increase starting at T$\sim$50K, as in previous studies \cite{Jiang2013, Cooley1995, Allen1979, Menth1969}, with saturation observed for T$<$5K.
The negative Hall resistivity in the saturation regime points towards electrons as the main carrier channel \cite{Allen1979, Cooley1995, Nickerson1971}.
The low-energy electron diffraction (LEED) pattern of Figure 1b proves the highly-ordered, long-range single-crystallinity of the cleavage surface, with sharp 1$\times$1 spots evident over a large incident electron energy range (60-300 eV).
This LEED pattern - observed for all cleaves of our crystals - is characteristic of high-quality SmB$_{6}$(001) surfaces, free from complicating surface reconstructions \cite{Aono1978} whose effect in \textit{k}-space is to obscure the fine details of the near-$E_{\textmd{F}}$ electronic band structure \cite{Miyazaki2012, Xu2013} which are so crucial to the physics at hand.
Figure 1c shows a schematic of the SmB$_{6}$ Brillouin zone, and its two-dimensional (2D) projection onto the
(001) plane.

Figure 1d shows the backdrop for the remainder of the action presented in this paper, in the form of a complete $E$($k_{\textmd{x}}$,$k_{\textmd{y}}$) dataset recorded from SmB$_{6}$.
These data have been acquired with $h\nu=70$ eV, corresponding to a $k_{\textmd{z}}$ value of 6$\pi$/a (see Figure S1), meaning the high symmetry directions probed in the \textit{bulk} Brillouin zone would be $\Gamma \textmd{X}$ and $\textmd{XM}$, as in the lower, green plane in Figure 1c.

In Figure 1d, three essentially non-dispersive bands can be easily identified at binding
energies ($E_{\textmd{B}}$) of 40, 170 and 960 meV.
These originate from the Sm 4$f$ states, attributed to the $^{6}$H$_{5/2}$, $^{6}$H$_{7/2}$
and $^{6}$F final state multiplets of the 4$f^{6}$$\rightarrow$4$f^{5}$ transitions, respectively \cite{Miyazaki2012, Neupane2013, Souma2002, Denlinger2000}.
In addition, these flat bands are crossed by highly-dispersive electronic states.
First-principles calculations predict the latter to have dominant Sm 5$d$ character, with mixed $d$-4$f$ character in energy regions (we refer to later as hybridization zones), in which there is a strong interaction with the 4$f$ states \cite{Lu2013, Neupane2013, Alexandrov2013}.

Already in the data-block of Figure 1d, these light 5$d$-related states can be seen to form elliptical constant energy contours when not in the hybridization zones. 
Figure 2a-2e shows how these elliptical contours grow steadily in size as $E_{\textmd{B}}$ is reduced, leading to an electron pocket centered on the $\overline{\textmd{X}}$ points of the (001)-projected Brillouin zone covering an area of 33\% of the latter for $E=E_{\textmd{F}}$.
These elliptical-contour electronic states are the first important group of states that cross $E_{\textmd{F}}$ in SmB$_{6}$ and will be referred to as the Ô$\textmd{X}$-statesÕ in the following.   

In panels l to o of Figure 2, a zoom around the bulk $\textmd{X}$-point illustrates the evolution of the $\textmd{X}$-states.
An elliptical guide to the eye has been added on top of the data for $E_{\textmd{B}}=430$ meV, and this same guide has been added to panels l,m and o.
The growth of the elliptical $\textmd{X}$-state from 1311 meV (Figure 2o) to 430 meV (Figure 2n) is evident, and for 77 meV binding energy (panel m), the elliptical intensity contour in the data is, as expected, larger than the $E_{\textmd{B}}=430$ meV schematic dashed ellipse. 
Interestingly, the data for 10 meV below the Fermi level show the $\textmd{X}$-state ellipse to be of the same size as $E_{\textmd{B}}=$430 meV.           

Figs. 2f-h provide the explanation for this renormalization of the $\textmd{X}$-state ellipses: in the hybridization zones - when the 5$d$-derived states cross the flat 4$f$-bands - the momentum distribution of the electronic states is altered radically. 
In Figure 2i, a schematic of the new energy contour is overlaid in yellow on the $E_{\textmd{B}}=40$ meV data, and panel j of Figure 2 shows that the identical contour (now in blue) perfectly describes the experimental intensity distribution  at $E_{\textmd{B}}=170$ meV.
Comparing the hybridization zone energy contours and the $\textmd{X}$-state ellipses as is done in Figure 2k, reveals the essentially four-fold symmetric fingerprint of the $d$-$f$ hybridization in \textit{k$_{x,y}$}-space, which our data pick up in both the 1st and 2nd Brillouin zones.
We point out that in the hybridization zones themselves, the experimental intensity is high at the center of the Brillouin zone (see Figs. 2f and 2g).
This, and the equivalence of the \textit{k}-space fingerprint of the hybridization at high energies (170 meV) and low energies (40 meV) is an important point, and will be returned to later when discussing the electronic structure at and close to the Fermi level in more detail. 

Closing our discussion of Figure 2, we note that a consequence of the $d$-$f$ hybridization is that for $E_{\textmd{B}}$ a little lower than the hybridization zone (e.g. $E_{\textmd{B}}=10$ meV in Figure 2l), although the elliptical shape of the $\textmd{X}$-states is reasserting itself, the $k$-value for these 5$d$-related states is reduced, compared to before the hybridization zone (see also the data-movies in Figs. S2, S2a and S2b of the supplementary information).
These phenomena - both in the hybridization zones as well as in the interaction-free energy windows - are in full agreement with data from bulk LDA, GGA and LDA+Gutzwiller calculations all of which predict states of mixed $d$-$f$ character upon interaction of the Sm 4$f$ and Sm 5$d$ states \cite{Lu2013, Alexandrov2013, Neupane2013}.
Realizing this brings with it the inescapable conclusion that the altered \textit{k}-space distribution and the discontinuity in the $\textmd{X}$-state $k$-values in the hybridization zones is nothing other than the signature of bulk $d$-$f$ hybridization.
Occam's razor excludes an explanation going beyond the bulk physics of the hybridization itself, meaning the $\textmd{X}$-state $k$-renormalization cannot be interpreted as a consequence of band inversion and the occurrence of topological protected electronic edge states.

\begin{figure*}
  \centering
  \includegraphics[width = 17 cm]{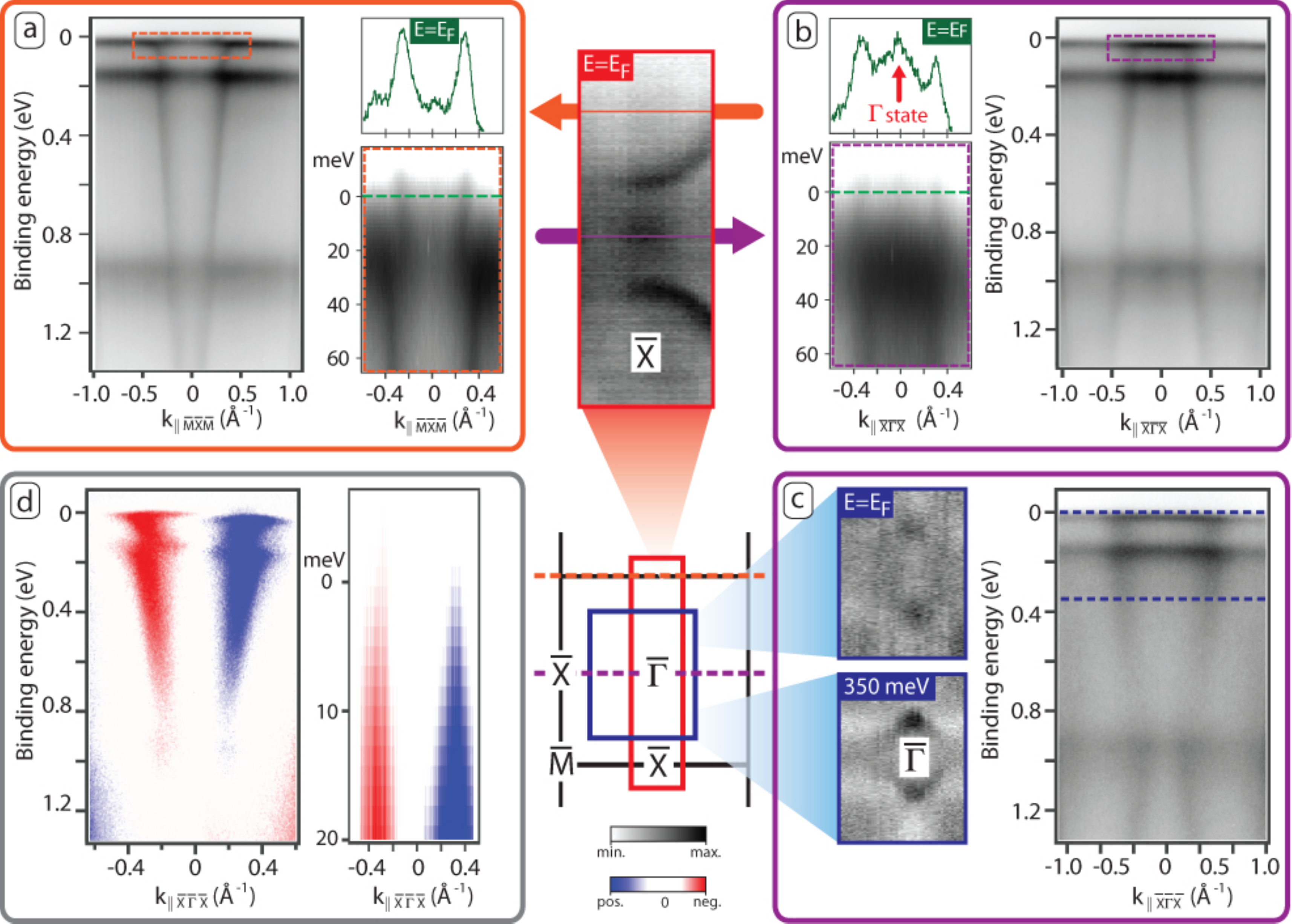}
  \caption{
\textbf{Energy dispersion and dichroic signature of the $\textmd{X}$-state; identification of an additional, $\Gamma$-state feature}
(a) and (b): data recorded for $k_{\textmd{z}}=6\pi$/a  ($h\nu=$70 eV) for the \textit{k}-space cuts along the $\overline{\textmd{MXM}}$ and $\overline{\textmd{X}\Gamma\textmd{X}}$ high symmetry directions as indicated by the color-coded arrows underlying the Fermi surface map in between the panels (integrated over $\pm$5 meV).
In each case a zoom of the dashed region is included, as is a momentum distribution line-cut in green for $E=E_{\textmd{F}}$.
In panel (a) these clearly show the $\textmd{X}$-states crossing $E_{\textmd{F}}$, and the same states seen as the outer two features in  (b) are joined by an enigmatic, third feature centered on the zone center dubbed the $\Gamma$-state. 
Panel (c) shows data along the same direction as that in panel (b), but now recorded for $k_{\textmd{z}}=4.5\pi$/a ($h\nu=$34 eV).
It is evident that these Sm 5$d$-dominated states now have an inverted dispersion with respect to panel (b). Plus, both the Fermi surface and the 350 meV energy contour (both integrated over $\pm$10 meV) possess significant intensity around the $\overline{\Gamma}$-point, rather than centered on the $\overline{\textmd{X}}$-point.
A schematic of the surface Brillouin zone is added between panels (c) and (d) showing, color-coded, the \textit{k}-space coverage of the constant energy surfaces and the \textit{k}-space trajectories giving the E(\textit{k}) images as dashed lines.
Finally, panel (d) illustrates that the dichroic signature of the Sm 5$d$-related bands, shown for different binding energy regimes and $k_{\textmd{z}}$ values.
Plotted is the difference in ARPES intensity between data recorded using left or right circular light.
It is clear that this dichroic signal has the same sign at all energies below 1 eV. In particular, the fact that the same dichroic response is seen for high binding energy (left panel, $h\nu=34$ eV, $k_{\textmd{z}}=4.5\pi$/a) and for the 20 meV closest to the Fermi level (right panel, $h\nu=21$ eV, $k_{\textmd{z}}=2\pi$/a) shows that the states giving rise to the elliptical Fermi surfaces around the $\overline{\textmd{X}}$ points are no different to those making up the large scale 5$d$-related dispersive features pointed out in Figure 1d.
}
\label{SmB6_Fig3_v5}
\end{figure*}

We now turn to Figure 3a, in which an $E$($k$) image for the ${\overline{\textmd{MXM}}}$ direction, a zoom and  the momentum distribution curve for $E=E_{\textmd{F}}$ are shown for $h\nu=70$ eV data, with the analogous cut along $\overline{\textmd{X}\Gamma\textmd{X}}$ shown in panel b. 
The location of these cuts in \textit{k}-space is indicated with the orange/purple arrows underlying the raw-data Fermi surface map shown between panels (a) and (b).

At variance with theoretical calculations predicting a Kondo hybridization gap at the Fermi level, the data of Figure 3 show the sample to have a robustly metallic character.
The zoom in panel a clearly shows the elliptical $\textmd{X}$-states crossing $E=E_{\textmd{F}}$, states which are also clearly identifiable in the data of both Figure 1d and Figure 2a.
In Figure 3b, however, the $E=E_{\textmd{F}}$ MDC clearly picks up a third maximum, located at $k=0$, in between the $\textmd{X}$-state peaks, indicating the existence of a second type of metallic state in SmB$_{6}$, which we forthwith refer to as the $\Gamma$-state.
We will return to the low energy dispersions visible in the zoomed panels of Figure 3a and b in a moment, but first discuss the data shown in Figs. 3c and d.

The data of in Figure 3c have been acquired with the same sample orientation as those of panel b, but with $h\nu=34$ eV.
This gives a $k_{\textmd{z}}$ value of 4.5$\pi$/a (see Figure S1). 
If there were no dependence of the $E$($k_{\textmd{x}}$,$k_{\textmd{y}}$) values of the Sm 5$d$-dominated states on $k_{\textmd{z}}$, as would be the case for a surface state, the dispersion relations shown in Figs. 3b and 3c should be indistinguishable, which is obviously not the case.

Figure 3c also shows two constant energy surfaces (upper at $E_{\textmd{F}}$, lower at 350 meV).
Both reinforce the conclusion that the electronic states - now also those at and close to the Fermi level - for a $k_{\textmd{z}}$ value of 4.5$\pi$/a do not form elliptical contours around the $\overline{\textmd{X}}$ points as for data acquired at $k_{\textmd{z}}=6\pi$/a ($h\nu=$70 eV: Figs. 1d, 2, and 3a) and $k_{\textmd{z}}=4\pi$/a ($h\nu=21$ eV: see Figure S1).
For $k_{\textmd{z}}=4.5\pi$/a, the momentum distributions are, in fact, centered around $\overline{\Gamma}$, rather than around $\overline{\textmd{X}}$. After a careful inspection of the
constant energy panels of Fig. 2 and Fig. 3c, one can readily conclude that - on one hand - the band topography at $E_{\textmd{F}}$ changes substantially with $k_{\textmd{z}}$, while - on the other hand - there are no major changes between the momentum distributions of the near-$E_{\textmd{F}}$ states and those of the Sm 5$d$ bulk bands acquired at higher-$E_{\textmd{b}}$ and the same $k_{\textmd{z}}$. Therefore, by presenting the complete band topography, we can confirm both the non-negligible bulk dispersion of the metallic features and their Sm 5$d$ origin .  

Turning now to Figure 3d, we offer a final argument for the essentially bulk, Sm 5$d$ character of the $\textmd{X}$-states both at higher energy and near $E_{\textmd{F}}$.
This panel uses data acquired along $\overline{\textmd{MXM}}$ using circularly polarized light, in which the difference of the photoemission signals acquired with circular left and circular right polarization is shown. 
The left-hand image ($h\nu=34$ eV) shows a clear dichroism in intensity \cite{Jiang2013} between the left and right branches of the Sm 5$d$-related, strongly dispersive feature already described in the context of panel (c) of the same figure.
The right-hand image is a zoom-in, showing only the first 20 meV ($h\nu=21$ eV, $k_{\textmd{z}}=4\pi$/a).
These data demonstrate identical dichroism behavior for the high energy, Sm 5$d$-derived states (for example for $E_{\textmd{B}}=430$ meV [see Figure 2n]) and the $\textmd{X}$-states within 20 meV of $E_{\textmd{F}}$.
This fact would suggest that they possess the same or very similar orbital character, as the polarization dependence of the photoemission matrix element is sensitive to the symmetry of the initial (and final) states involved \cite{Westphal1989,Kaminski2002,Vyalikh2008}.
This interpretation is backed up by the lack of dichroism at $E$($k_{\textmd{x}}$,$k_{\textmd{y}}$) locations where the flat bands have essentially pure Sm 4$f$ character - i.e. where the bands strongly differ in orbital character from the $\textmd{X}$-states. 

The electronic states relevant for connection to the transport data  on SmB$_{6}$ which is presently creating such excitement are - of course - those very close the Fermi level.
The zoomed $E$($k$) images for the $\overline{\textmd{MXM}}$ and $\overline{\textmd{X}\Gamma\textmd{X}}$ directions in Figs. 3a and 3b directly address the states very close to $E_{\textmd{F}}$.
In both cases, there is a swathe of high intensity between binding energies of 40 and 20 meV, related to the flat bands connected to the Sm 4$f$\hspace{0.7mm}$^{6}$H$_{5/2}$ state.
The Sm 5$d$-related $\textmd{X}$-states are visible entering the high intensity strip from below and leaving it approaching the Fermi level.
Comparing the $\textmd{X}$-state $k$-values at higher and lower $E_{\textmd{B}}$, the $k$-discontinuities in the $\textmd{X}$-states are very evident, signaling the strong $d$-$f$ hybridization, which in turn is consistent with the attribution of the high intensity strip between 20 and 40 meV binding energy to the 4$f$ states.

These same zoomed images, although a powerful argument for $d$-$f$ hybridization, also make it very clear that our experiment is not able to resolve the hybridization gaps which must accompany the $k$-renormalization phenomenon.  
What these data can pin down without any doubt, is the fact that the chemical potential in the (near-surface) region probed by ARPES from a super-structure-free, (001) terminated, high quality SmB$_{6}$ crystal is not in the Sm 5$d$-Sm 4$f$\hspace{0.7mm}$^{6}$H$_{5/2}$ hybridization gap, but that this hybridization zone lies some 20 meV below the Fermi level.
This reveals an overestimation of the hybridization energy gaps in LDA calculations \cite{Lu2013, Alexandrov2013}.

To put the similarities and differences between our experimental data and typical theoretical predictions into sharp focus, in Figures 4a and 4b we illustrate the changes that can be expected to the dispersion relation of the Sm 5$d$ and 4$f$ states near $E_{\textmd{F}}$ for the two high symmetry directions probed experimentally in Figures 3a and 3b, when the hybridization interaction is turned off or on.
In each panel, the right-hand part (positive $k$-scales) illustrates the non-hybridized situation using solid, purple lines. 
Using the \textit{bulk} LDA+Gutzwiller theory results of Lu \textit{et al.} \cite{Lu2013} as a basis, we sketch the steep Sm 5$d$ band, and three different, much flatter 4$f$ bands.
In the left-hand parts (negative $k$	-scales), the solid green ($\overline{\textmd{MXM}}$) and orange ($\overline{\textmd{X}\Gamma\textmd{X}}$) lines show the situation after the hybridization is switched on \cite{Lu2013}.
The hybridized bands are copied into the right-hand parts of both Figure 4a and 4b, making it clear that the hybridization gaps open up where the 5$d$ and 4$f$ bands of identical symmetry ($\Gamma_{7}$ \cite{Lu2013}) cross one another.

In the lowest two panels - Figure 4c1 and 4d - the same illustrative theory curves (hybridized case only) are superimposed on the relevant ARPES $E$($k_{\textmd{x}}$,$k_{\textmd{y}}$) images.
Two modifications have been made to the theory curves in Figure 4 compared to those in \cite{Lu2013}: the theory has been shifted 20 meV to higher binding energy to account for the position of the chemical potential in the ARPES experiment and the total bandwidth of the $^{6}$H$_{5/2}$ 4$f$ multiplet has been re-scaled to match the experimental data.

What is clear from the comparison shown in Figs. 4c1 and 4d, is that after using the Sm 4$f$ states to pin the energy scale of the theory, the rest of strongly dispersive $d$-$f$ hybridized bulk bands (that have dominant $d$ character), fall nicely on top of the experimental features without any adjustment, confirming our assignment of the $\textmd{X}$-states made earlier on the basis of the $k$-discontinuity in their electronic dispersion relation.

Inspecting the correctly energy-scaled theory curves shown in Figure 4, it is immediately clear that the hybridization gaps within which possible topologically protected edge states could be situated would be (a) of the order 20 meV below the Fermi level for the experimentally determined chemical potential and (b) would be only 5-10 meV along both the $\overline{\textmd{XM}}$ and $\overline{\Gamma\textmd{X}}$ high-symmetry directions. 
Resolving such very small hybridization gaps between the 5$d$ and 4$f$-derived levels so as to be able to identify a massless Dirac cone as an in-gap state is simply beyond the experimental resolution of the present or recent ARPES studies on SmB$_{6}$ \cite{Miyazaki2012, Neupane2013, Xu2013, Jiang2013}.

\begin{figure}
  \centering
  \includegraphics[width = 7.3 cm]{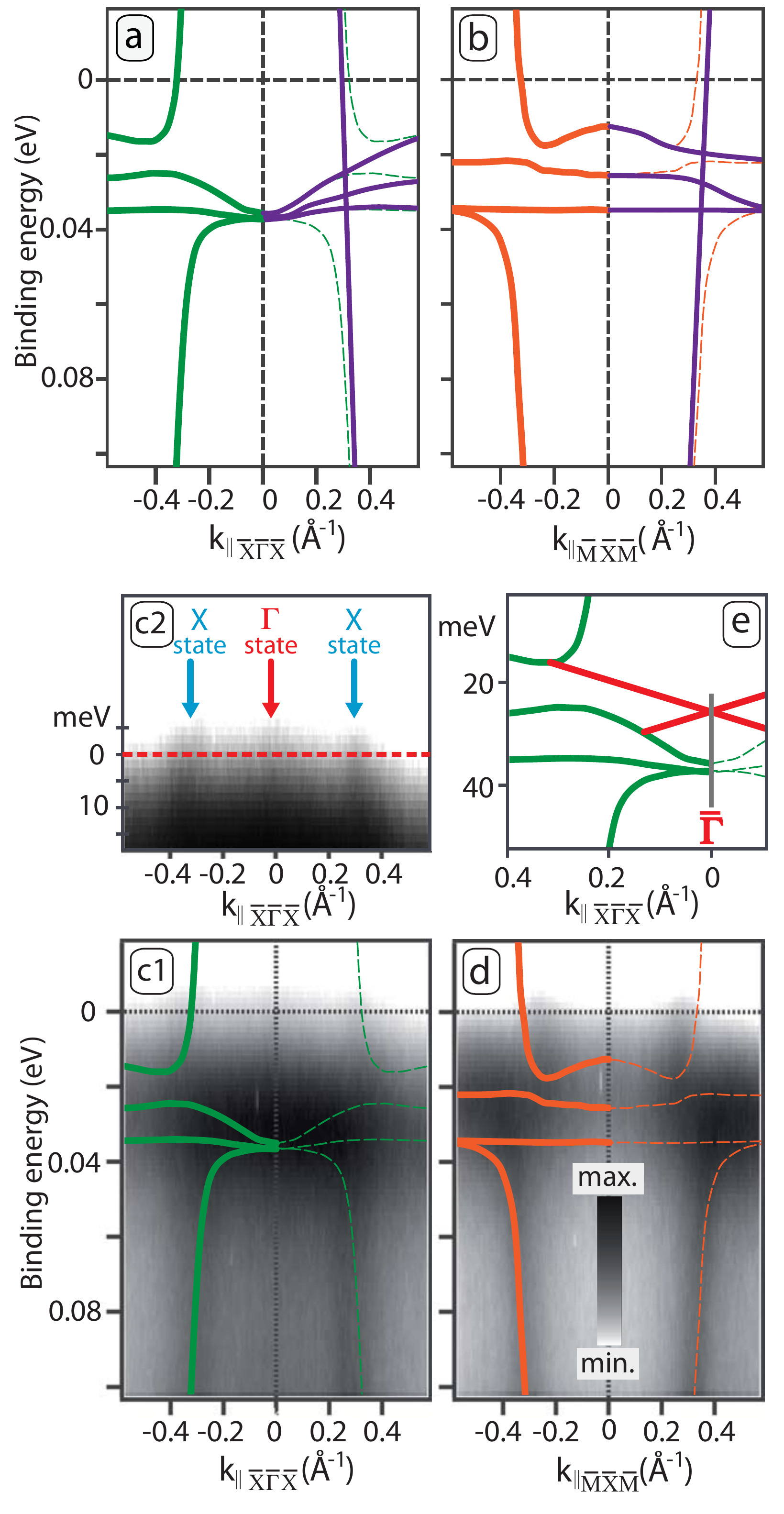}
  \caption{
\textbf{$\textit{k}$-discontinuities, position of the chemical potential and hybridization renormalization from comparison with bulk, \textit{ab initio} theory} 
(a) and (b): Full lines in the left(right) parts illustrate hybridised(non-hybridised) $d$ and $f$ bands along two high symmetry directions in \textit{k}-space. The lines are adaptations of the bulk LDA+Gutzwiller theory data of Ref. \cite{Lu2013}. The hybridised cases are overlaid as thin dashed lines in the right parts to aid identification of the hybridization gapping.
(c1) and (d): the same theory curves are now overlaid on the experimental band structure along $\overline{\textmd{X}\Gamma\textmd{X}}$ and $\overline{\textmd{MXM}}$. $k_{\textmd{z}}=6\pi$/a ($h\nu=70$ eV).
A very good match is obtained when the total bandwidth of the theoretical 4$f$ multiplet is renormalised by a factor of one third compared to the original theory data \cite{Lu2013}, with a simultaneous rigid shift of the theoretical data 20 meV to higher binding energy, carried out to account for a difference in chemical potential between theory and experiment.
This renormalization means the 5$d$-4$f$ hybridization gaps are only 5-10 meV.
From the panels (c1) and (d), it is clear that the $\textmd{X}$-states, including the \textit{k}-discontinuity which is evident on comparing below and above the hybridization zone, are described very well by the energy-scaled \textit{bulk} theory, removing them from the suspect line-up as far as topological surface states go. In contrast, the $\Gamma$-states, which are highlighted in the zoomed portion of the $\overline{\textmd{X}\Gamma\textmd{X}}$ data in panel (c2) are missing in the bulk theory. This is a signal of their promise in terms of identifying potential topological surface states in future ultrahigh resolution ARPES experiments. Panel (e) shows a zoom of the theory around $\overline{\Gamma}$, together with a sketch of a massless Dirac cone centered at the $\overline{\Gamma}$-point as a guide for future experiments.}
\label{SmB6_Fig4_v5}
\end{figure}

The comparison shown in Figures 4c1 and 4d is able to explain all the observed features in the experimental data with the results of \textit{bulk} hybridization calculations with the exception of what we called the $\Gamma$-state, most clearly seen at the Fermi level for $\textit{k}=0$ in the data of Figure 4c1 and the related zoom in panel (c2).
This state has not been predicted in bulk calculations \cite{Lu2013, Alexandrov2013, Neupane2013}.

What is the profile of this as yet unexplained $\Gamma$-state? 
Firstly, as mentioned in the discussion of the data of Figure 3b and 3c, it shows up as increased intensity centered at and near the zone center.
The data presented in Figures 2f and 2g show that this zone center intensity is even more pronounced for energies in the middle of the hybridization zones.
This would strongly suggest that the $\Gamma$-state is related to 5$d$-4$f$ hybridization, and therefore to the opening of Kondo-derived energy gaps.

Interestingly, slab calculations carried out within an LDA+Gutzwiller approach \cite{Lu2013} have predicted surface states of topological character within these gaps both at $\overline{\textmd{X}}$ and $\overline{\Gamma}$.
In Figure 4e we provide a sketch, showing a zoomed portion of the theory curves for the hybridized states around $\overline{\Gamma}$ (in green) and a Dirac cone at the zone center.
Consideration of Figure 4c2 and 4e together would suggest that the $\Gamma$-state would be the most promising candidate for an experimental indication for the possible existence of topological states. As expected, the ARPES intensity from the $\Gamma$-state fades away as one moves away from the hybridization zones (Fig. 2). However, it is still non-negligible at $E_{\textmd{F}}$, an issue which could be explained by the superposition of electronic band structures with a different chemical potential coming from a dominant (i.e. 1$\times$1) and a minority atomic arrangement \cite{Hoffman2013}. Despite the high level of clarity of the ARPES data from SmB$_{6}$ presented here, the weakness and associated indistinctness of the $\Gamma$-state in either our or others' ARPES data made public to date precludes an unambiguous determination of the linear $E$($k_{\textmd{x}}$,$k_{\textmd{y}}$) relation for this state and whether it is independent of $E$($k_{\textmd{z}}$). These observations would form the first steps towards an ARPES-based conclusion on whether SmB$_{6}$ is a TKI.

We now take a step back from the microscopics of the ARPES data and the very strenuous conditions that will need to be met in future ARPES experiments on SmB$_{6}$ in order to positively identify topological edge states. 
We close by summarizing what has been discovered from the data presented here, and then discuss how that fits into the larger picture of SmB$_{6}$.  

Our data have established the presence of two different types of states crossing the Fermi level in unreconstructed, high-quality crystals of SmB$_{6}$.

The first, the $\textmd{X}$-states, possess all the characteristics of bulk Sm 5$d$-derived bands, forming elliptical constant energy contours for energies away from the 5$d$-4$f$ hybridization zones. 
This assignment is based upon the following facts:

a) The $\textmd{X}$-states exhibit clear sensitivity to $k_{\textmd{z}}$, belying their significant 3D character. 

b) The experimentally determined Fermi surface of SmB$_{6}$ does not lie in such a hybridization zone, but 20 meV above the energy of the 4$f^{6}$H$_{5/2}$ multiplet, and thus is dominated by the elliptical contours of the $\textmd{X}$-states. In addition the dichroic signature of the low-lying $\textmd{X}$-states is the same as that for the 5$d$-band at energies well away from both $E_{\textmd{F}}$ and hybridization zones.   

c) The k-space signature of the lowest lying 5$d$-4$f^{6}$H$_{5/2}$ hybridization zone within 20 meV of $E_{\textmd{F}}$ is essentially identical to that of the next hybridization zone involving the 4$f^{6}$H$_{7/2}$ level at a binding energy of 170 meV.   

The conclusion just made that the $\textmd{X}$-states which cross $E_{\textmd{F}}$ are not Ôin-gapÕ states is in line with the temperature-dependent ARPES data shown in Refs. \onlinecite{Jiang2013} and \onlinecite{Xu2013} where what we have identified as the $\textmd{X}$-states and Sm 5$d$ bands merge when T is higher than $\sim$100K. In keeping with experimental determination of Sm valence fluctuations \cite{Mizumaki2009} and spectroscopic data \cite{Zhang2013, Souma2002, Nozawa2002}, this temperature value is within the regime which marks the onset of appreciable Kondo hybridization.
The same T-dependent ARPES data \cite{Xu2013, Jiang2013} also suggests that no significant changes happen in the electronic bandstructure as T decreases from the 38K used in this study to 10K, as both temperatures are well into the Kondo regime \cite{Zhang2013, Mizumaki2009, Souma2002, Nozawa2002, Cooley1995}.

The second type of state we have clearly identified is the $\Gamma$-state: located at or close to the center of the Brillouin zone.
These states are maximal in intensity in the hybridization zones between the 5$d$ and 4$f$ states, and are missing from all theory calculations for the bulk of SmB$_{6}$, making them candidates for possible topological edge states.

Over and above the clear identification of these two main groups of frontier electronic states, our ARPES data has unambiguously proven that:

i) the predicted hybridization gaps of 10-15 meV \cite{Lu2013, Neupane2013} are, in fact, overestimations, rather than 
underestimations, as stated in previous ARPES studies \cite{Neupane2013, Xu2013, Jiang2013, Miyazaki2012, Souma2002, Nozawa2002}.  We estimate the true hybridization gaps in the 5$d$-4$f^{6}$H$_{5/2}$ hybridization zone to be significantly less than 10 meV.

ii) the experimentally determined location of the chemical potential for the states near the (001) surface is some 20 meV above the lowest lying 5$d$-4$f^{6}$H$_{5/2}$ hybridization zone, rather than as theory predicts centered within it.

Theory has predicted that the topological surface states of cubic TKIs such as SmB$_{6}$ should be characterized by heavy
effective masses \cite{Alexandrov2013}.
This is in stark contrast to both the light, graphene-like effective masses we observe for the $\textmd{X}$-states (and what we see of the $\Gamma$-state) as well with recent magnetotransport data, which report two distinct Fermi surface signals with light effective masses as candidates for the topological surface states (TSS) \cite{Li2013}.
The magnetotransport frequencies correspond to Fermi contour areas of 0.2\% ($\alpha$-state) and 1.23\% ($\beta$-state) of the Brillouin zone, which are clearly smaller than the $\textmd{X}$-state electron pockets, which we survey to cover 33\% of the Brillouin zone.
In addition, the clearest of the two resonances (the $\beta$-state) is stated to originate from the (101) surface of the SmB$_{6}$ crystal \cite{Li2013}.
The speculation that the $\alpha$-state seen in magnetotransport is related to the (001) plane \cite{Li2013} is at least in line with our finding that the hybridization gaps in which such a TSS would live at the (001) surface are only 5-10 meV: a high Fermi velocity Dirac cone squeezed into a gap of 5 meV would have to have a very small maximal cross section.

Strong electron correlations have been previously reported to be at the origin of a strong renormalization of the LDA-predicted Sm 4$f$ bandwidth with respect
to experimental observations \cite{Miyazaki2012}. In line with these results, a later study proposed that LDA+Gutzwiller calculations, a theoretical tool developed for
correlated electron systems, can better describe the electronic band structure of SmB$_{6}$ \cite{Lu2013}. We have chosen this theoretical data for comparison with the experimental band structure (Fig. 4), but, as already mentioned, a further renormalization is unavoidable to obtain a satisfactory agreement with the experiment. A comparison of theoretical calculations and ARPES results on similar divalent hexaborides has also identified several discrepancies on gap 
sizes and bandwidths \cite{Denlinger2002}. Moreover, for the same compounds, it is common that the experimental position of the surface chemical potential lies in the cation $d$-derived conduction band instead of in the bandgap between the latter and the boron $p$ bands as proposed theoretically \cite{Denlinger2002}.
As a result, electron pockets forming elliptical contours around $\textmd{X}$ have been reported for EuB$_{6}$ and SrB$_{6}$, very
similar to the Fermi surface $\textmd{X}$-state contours shown in Figures 2 and 3 of this work and in other recent ARPES studies of SmB$_{6}$ \cite{Neupane2013, Jiang2013, Xu2013}.
Chemical potential shifts in divalent hexaborides have been attributed to the formation of boron vacancies, with the area of the $\overline{\textmd{X}}$ contours being related to their density at the surface \cite{Denlinger2002}.
In the case of SmB$_{6}$, the effect of stoichiometry on the energy position of the chemical potential has been investigated in an early transport study, where it was concluded that defects such as Sm vacancies could shift the Fermi energy away from the gap value
\cite{Allen1979}.
Therefore the discrepancies between theoretical predictions and the ARPES results presented here on the hybridization gap magnitude and the position of the chemical potential fit with other hexaboride data and provide a natural explanation of the ARPES results on SmB$_{6}$(001). \\

\section*{Concluding Remarks}

In the preceding discussion, key facts have been established as regards the character and origin of the low lying electronic states observed in ARPES from UHV-cleaved (001) surfaces of SmB$_{6}$. We now return to the observation at the core of this research field, namely the unconventional behavior of the resistivity vs. temperature for SmB$_{6}$, which is accepted to be a bulk Kondo insulator.

The transport data on cooling below 100K show increasing resistance, valence fluctuations are seen by bulk spectroscopic probes \cite{Mizumaki2009} and both our (Fig. 2) and other \cite{Denlinger2000, Jiang2013} ARPES data possess the tell-tale signs of $d$-$f$ hybridization. However, the ARPES data reported here clearly also show the X-states at the surface of UHV-cleaved SmB$_{6}$(001) to be metallic, thereby creating a dichotomy with the gapped Kondo insulating state in the bulk.
One resolution is to suggest that the X-states are topological surface states, living in the Kondo hybridization gap \cite{Xu2013, Jiang2013, Neupane2013}. Our ARPES data render this explanation untenable, in keeping with the Fermi surface areas extracted from recent quantum oscillation data \cite{Li2013}.
The key here is the realization that the chemical potential sensed in an ARPES experiment Ð i.e. that at the surface and the near-surface region Ð could be different to that in the bulk, Kondo insulating state.
Our ARPES data indeed show the Fermi level to be $\sim20$ meV above the lowest lying $d$-$f$ hybridization gap, whereas in the bulk Kondo insulator $E_{\textmd{F}}$ is within the gap. 
The picture sketched above would result in a two-conduction channel scenario (one bulk, one surface), which could explain the observed $T$-dependence of the resistivity \cite{Zhang2013}.

Naturally, the question now arises as to whether there is room at all for the topologically non-trivial surface states consistently predicted by theory for stoichiometric SmB$_{6}$?
We answer with reference to the $\Gamma$-states, seen clearly in our ARPES data in the Fermi surface map of Fig. 3 and also in Fig. 4c2. These states are strongest in the hybridization gaps and are centered at $\overline{\Gamma}$, just as predicted from theory for the topological surface states \cite{Lu2013, Alexandrov2013}.
Thus, were the surface of SmB$_{6}$ to share the same Ð mid-gap - chemical potential as the bulk, then at sufficiently low temperature the  $\Gamma$-states would be the only metallic states, and could cause the saturation of the resistivity, possibly exhibiting hallmarks of topological character in magnetotransport  experiments \cite{Li2013}.
An ARPES proof that the $\Gamma$-states of SmB$_{6}$ (001) are indeed topological in nature will require the resolution of a massless dispersion relation and spin-momentum locking for a Dirac cone of states squeezed into a hybridization gap of order 5 meV. This is beyond the reach of current ARPES instrumentation. 

After completion of this paper, a new ARPES + LDA study agreed with our conclusion of the essentially bulk nature of the X-states in UHV-cleaved SmB$_{6}$(001) \cite{Zhu2013}. In addition, the discussion of polarity-induced effects \cite{Hoffman2013,Zhu2013}, also coupled to chemical potential shifts \cite{Hoffman2013} is consistent with the scenario from our ARPES data sketched above.\\

\section*{Methods}

\subsection*{Sample growth and preparation of clean surfaces}SmB$_{6}$ single crystals were grown in an optical floating zone furnace (Crystal Systems Inc. FZ-T-12000-S-BU-PC) under 5 MPa pressure of high purity argon gas \cite{Bao2013}. The growth rate was 20 mm per hour with the feed and seed rods counter rotating at 30 rpm. Samples were cleaved at 38K at a pressure lower than 3.0$\times$10$^{-10}$ mbar.
\subsection*{Angle resolved photoelectron spectroscopy}
ARPES Experiments were performed at the UE112-PGM-2a-1$^{2}$ beamline (BESSY II storage ring at the Helmholtz Zentrum Berlin) using a Scienta R8000 hemispherical electron analyzer and a six-axis manipulator. The pressure during measurements was 1.0$\times$10$^{-10}$ mbar, while the sample temperature was maintained at 38K, well into the Kondo
regime \cite{Zhang2013, Mizumaki2009, Souma2002, Nozawa2002, Cooley1995}. The energy position of the Fermi level was determined for every cleave by evaporation of Au films onto the sample holder, in direct electrical contact with the crystal.
Measurements have been acquired with the [100] direction of the sample oriented along the analyzer
slit (i.e. $k_{\textmd{y}}$ in Figs. 2 and 3), while the whole Brillouin zone was spanned by sequential rotation of the polar
angle. The conversion of the photon energy to the relevant $k_{\textmd{z}}$ values was carried out assuming free electron final states
and an inner potential V$_{\textmd{o}}$ of 16 eV (supplementary Figure S1). Our $k_{\textmd{z}}$ values are in good agreement
with those of Refs. \onlinecite{Jiang2013, Miyazaki2012, Xu2013} and \onlinecite{Denlinger2000}.
\\
\section*{Acknowledgements}    
This work is part of the research programme of the Foundation for Fundamental Research on Matter (FOM), which is part of the Netherlands Organization for Scientific Research (NWO). In addition, we are grateful to F. Massee and M. Grioni for insightful comments.

\end{document}